\documentclass[12pt]{article}
\usepackage[dvips]{graphicx}
\usepackage{amsmath}
\usepackage{amssymb}

\usepackage{epsfig,amsmath,amssymb,verbatim,mathrsfs}
\def\beq{\begin{eqnarray}}
\def\eeq{\end{eqnarray}}
\def\bea{\begin{eqnarray}}
\def\eea{\end{eqnarray}}
\def\blfootnote{\xdef\@thefnmark{}\@footnotetext}

\newcommand{\ir}{\text{\tiny IR}}
\newcommand{\uv}{\text{\tiny UV}}

\newcommand{\be}{\begin{equation}}
\newcommand{\ee}{\end{equation}}

\renewcommand{\thefootnote}{\roman{footnote}}

\begin{document}
\begin{titlepage}
\noindent

\setlength{\baselineskip}{0.2in}
%\vspace{1cm}
\flushright{May 2015}

\vspace{1.2cm}
%\vspace{0.1cm}

\begin{center}
  \begin{Large}
    \begin{bf}
  Regarding the Radion in Randall-Sundrum Models with Brane Curvature
\end{bf}
  \end{Large}
\end{center}
\vspace{0.2cm}

\begin{center}
\begin{large}
{Barry~M.~Dillon,$^{1}$ Damien~P.~George$^{2,3}$ and Kristian~L.~McDonald$^{4}$}\\ %\footnote{Email: kristian.mcdonald@sydney.edu.au}}\\
\end{large}
\vspace{1cm}
  \begin{it}
$^1$ Department of Physics and Astronomy, University of Sussex,\\ Brighton,  BN1 9QH, U.K.\\
\vspace{0.1cm}
$^2$ DAMTP, CMS, University of Cambridge, Wilberforce Road,\\ Cambridge, CB3 OHA, U.K.\\
\vspace{0.1cm}
$^3$ Cavendish Laboratory, University of Cambridge, JJ Thomson Avenue, \\Cambridge, CB3 OHE,  U.K.\\
\vspace{0.1cm}
$^4$ ARC Centre of Excellence for Particle Physics at the Terascale,\\
School of Physics, The University of Sydney, NSW 2006, Australia
\vspace{0.1cm}
\end{it}
\vspace{0.5cm}

\end{center}

%\center{\today}

\begin{abstract}
In Randall-Sundrum models, one typically expects the radion to be the lightest new ``gravity" state, as it is dual to a composite pseudo-Goldstone boson associated with conformal symmetry breaking in the IR.  Here, we investigate the effects of  localized brane curvature on the properties of the radion in Goldberger-Wise stabilized Randall-Sundrum models. We point out that both the radion mass and coupling to brane  matter are sensitive to the brane curvature. Radion/Higgs kinetic mixing, via an IR-localized non-minimal coupling to the Higgs, is  also investigated, in relation to the ghost-like radion that can occur for $\mathcal{O}(10)$ values of the IR curvature (as required to significantly suppress the first Kaluza-Klein graviton mass). We also discuss a class of speculative IR localized terms involving the radion. Basic comments regarding the dual 4D theory are offered.
\end{abstract}

\vspace{2cm}

\end{titlepage}
%\pacs{PACS numbers: }
%]

\renewcommand{\thefootnote}{\arabic{footnote}}
\setcounter{footnote}{0}
\setcounter{page}{1}
%\setcounter{figure}{0}
%\setcounter{table}{0}

%\tableofcontents

\vfill\eject

%\end{comment}

%%%%%%%%%%%%%%%%%%%%%%%%%%%%%%%%%%%%%%%%%%%%%%%%%%%%%
\section{Introduction}
The Randall-Sundrum (RS) model provides a natural means by which to generate hierarchically separated, radiatively-stable mass scales~\cite{Randall:1999ee}. Accordingly, it has received much attention as  a candidate solution to the hierarchy problem.  The model employs a warped extra dimension, namely a gravitational background, with factorizable geometry, that is sourced by a bulk cosmological constant and non-trivial brane tensions. The use of localized branes in the 5D spacetime explicitly breaks the 5D diffeomorphism symmetry yet preserves the requisite 4D symmetry. Consequently the most-general Lagrangian for the model, consistent with the symmetries, allows localized 4D terms that break the 5D diffeomorphism symmetry.

Included among the set of such terms are the so-called ``brane curvature" terms, which can be thought of as localized 4D kinetic terms for the bulk graviton.  These terms have received some attention in the literature~\cite{Davoudiasl:2003zt}, though generally they are assumed subdominant. Nonetheless, they should appear in the most-general Lagrangian. Recently the brane curvature terms received attention in relation to the 750~GeV diphoton excess observed at the LHC. In particular, it was shown that large brane curvature (i.e.~with an $\mathcal{O}(10)$ dimensionless coefficient) can modify the spectrum of Kaluza-Klein (KK) gravitons in a non-trivial way, permitting  the lightest KK graviton to have a 750 GeV mass, while retaining an $\mathcal{O}(\mathrm{TeV})$ lightest KK vector~\cite{Falkowski:2016glr,Hewett:2016omf,Dillon:2016fgw}. 

In addition to the KK gravitons, the bulk 5D metric gives rise to a gravi-scalar fluctuation, known as the radion~\cite{Csaki:2000zn,Cheung:2000rw}. This field is massless unless the length of the extra dimension is stabilized. After stabilization it acquires a mass that is sensitive to the backreaction of the stabilizing dynamics. The best-studied method for stabilizing the extra dimension relies on a bulk scalar that develops a non-trivial background value to generate a potential for the radion (as proposed by Goldberger and Wise (GW)~\cite{Goldberger:1999uk,Goldberger:1999un}). 

The common origin of the radion and KK gravitons, as fluctuations of the bulk metric, means both are sensitive to brane curvature terms. Motivated by recent interest in large brane curvature, in this work we investigate some effects of brane curvature terms on the properties of the radion in a GW-stabilized RS model. We consider the modification to the radion mass and couplings due to the brane curvature, and further consider the effects of an IR localized non-minimal coupling to the Standard Model (SM) Higgs. Our results generalize a number of the corresponding  expressions in Ref.~\cite{Csaki:2000zn} to include the effects of brane curvature. We find that, in the GW stabilized model, a non-minimal coupling to the IR Higgs does not allow one to avoid the ghost-like radion that arises for $\mathcal{O}(10)$ values of the IR curvature. However, we further comment on some speculative IR localized terms that may help remove the ghost-like radion.

Before proceeding we note that a number of works  have considered the RS model in relation to the 750~GeV diphoton excess; see e.g.~Ref.~\cite{Cox:2015ckc}.  For additional discussion of the spin-2 explanation see Ref.~\cite{Sanz:2016auj}. In our analysis, we present the results for the RS model with a UV scale of $M_*\sim\mathcal{O}(M_{Pl})$. However, the results are readily adapted to the little RS model~\cite{Davoudiasl:2008hx}, and related warped models~\cite{McDonald:2010jm}, for which $M_*\ll M_{Pl}$.

The layout of this paper is as follows. In Section~\ref{sec:setup} we
describe the setup for our analysis and present some consequences of the brane curvature terms for the case of a massless radion (i.e., a non-stabilized RS model). These results prove useful for subsequent analysis. We explore the radion coupling to brane-localized matter in Section~\ref{sec:radion-matter-GI}, and turn to the more-general case of GW stabilized RS models in Section~\ref{sec:GW}. The effects of a non-minimal coupling with an IR localized SM Higgs are studied in Section~\ref{sec:radion_higgs_mixing} and additional IR-localized terms for the radion are considered in Section~\ref{sec:new_ir_rad}. Comments regarding the interpretation in the dual 4D theory are given in Section~\ref{sec:ads/cft} and we conclude in Section~\ref{sec:conc}.
%%%%%%%%%%%%%%%%%%%%%%%%%%%%%%%%%%%%%%%%
%%%%%%%%%%%%%%%%%%%%%%%%%%%%%%%%%%%%%%%%%%%%%%%%%%%%%
\section{The Randall-Sundrum Model with Brane Curvature\label{sec:setup}}
To study the effects
of the brane curvature terms on the metric fluctuations, we
 employ the interval approach to brane-world
gravity~\cite{Lalak:2001fd,Carena:2005gq,Bao:2005bv}. This approach enables a
transparent treatment of boundary curvature terms, which simply  modify the
boundary conditions (BCs) for metric fluctuations. However, one must be careful to correctly identify the available gauge freedoms in the presence of such terms (for detailed discussion see Ref.~\cite{George:2011sw}). Before
proceeding, we note that earlier works have considered the effects of
brane curvature terms in the RS framework using the orbifold
picture~\cite{Davoudiasl:2003zt,Davoudiasl:2005uu}, and for
$AdS_5/AdS_4$ in the
interval approach~\cite{Carena:2005gq,Bao:2005bv}. Let us also note that some content in the following sections has overlap with Ref.~\cite{George:2011sw}. We include it here so the presentation is coherent and  (relatively) self-contained, and note that ($i$) we present a number of extra results, in relation to IR curvature, that weren't given in Ref.~\cite{George:2011sw}, due to the focus on UV curvature in that work; ($ii$) in the current presentation we focus on the case with $M_*\sim M_{Pl}$, relevant for RS models, as opposed to the low UV-scale models of interest in Ref.~\cite{George:2011sw};  $(iii)$ we subsequently  generalize these results to include a non-minimal coupling with an IR Higgs and additional IR terms for the radion.

The RS model employs a warped extra
dimension, labeled by the
coordinate $y\in[0,L]$, with a UV (IR) brane of characteristic
energy  $M_*$ ($e^{-kL}M_*$) located at
$y=0$ ($y=L$). The metric has the form
\beq
ds^2 = e^{-2ky}\eta_{\mu\nu}dx^{\mu}dx^{\nu} +
dy^2= G_{MN} dx^{M}dx^{N},
\label{bulkmetric}
\eeq
where $M,N,..$ ($\mu,\nu,..$) are 5D (4D) Lorentz indices and
$k$ denotes the $AdS_5$ curvature. The corresponding action,
including brane localized curvature terms, is
\bea
S&=&\int_{\mathcal{M}} d^5x \sqrt{-G}\
\left\{2M_*^3\mathcal{R}-\Lambda\right\}\ +\ \sum_i\int d^4x \sqrt{-g_i}\left\{M_i^2 R_i- V_i/2\right\}\nonumber\\
& &\qquad\qquad+\ 4M_*^3\oint_{\partial \mathcal{M}} \sqrt{-g_i}\ K.\label{gravity_action}
\eea
The bulk Ricci scalar $\mathcal{R}$ is constructed with the bulk metric $G_{MN}$
and $M_*$ is the 5D gravity scale.  The brane localized curvature $R_i$ is constructed with the brane metric $g_{uv}^i$  (the restriction of $G_{\mu\nu}$ to the relevant boundary), and has coefficient $M_i$ on the $i$th boundary ($i=\uv,\,\ir$).
The last term is the usual Gibbons-Hawking boundary term~\cite{Gibbons:1976ue},
with $K$ being the extrinsic curvature of the manifold $\mathcal{M}$. This term
is included to obtain consistent Einstein equations on the interval~\cite{Lalak:2001fd}.
The action includes a bulk
cosmological constant $\Lambda$, and brane tensions $V_i$, which  take their usual RS values,
$V_i=-24kM_*^3\theta_i$, with bulk
curvature $k=\sqrt{-\Lambda/24M_*^3}$ and we use the notation $\theta_\uv=-\theta_\ir=-1$. For future purposes, we define the dimensionless brane curvature coefficients $v_i=M_i^2k/M_*^3$ and $w_i=V_i/2M_*^3k$.

The calculation of the effective 4D Planck mass   gives
\bea
M_{Pl}^2=\frac{M_*^3}{2k}\left\{1+v_\uv - (1-v_\ir)e^{-2kL} \right\}.\label{eq:4D_planck}
\eea
This includes contributions from both the bulk and
brane intrinsic curvatures. Observe that the Planck mass is rather insensitive to the IR curvature, while a constraint of $(1+v_\uv)>0$ is required to ensure positivity of the Planck mass (equivalently, to avoid a ghost-like massless graviton). The different pieces have distinct interpretations in the dual 4D picture, as we discuss in Section~\ref{sec:ads/cft}. Variation of the bulk action gives the standard (bulk) equations of motion,
\bea
\mathcal{R}_{MN}-\frac{1}{2}G_{MN}\mathcal{R}=-\frac{\Lambda}{4M_*^3}G_{MN}.
\eea
The boundary conditions follow from the variations of the
4D brane action and the Gibbons-Hawking term, combined with surface terms resulting
from the variation of the bulk action, giving~\cite{Carena:2005gq}
\bea
\left[\frac{v_i}{k}
  \left(R_{\mu\nu}-\frac{1}{2}g_{\mu\nu}R\right)+\frac{1}{2}g_{\mu\nu}kw_i+\theta_i\sqrt{G^{55}}(g_{\mu\nu,5}-g_{\mu\nu}\
  g_{\alpha\beta,5}\
  g^{\alpha\beta})\right]_{y=y_i}=0.
\eea
This equation should be evaluated
separately at the boundaries $y=0,L$.
We work in a
``straight gauge," defined by
$G_{\mu5}=0$~\cite{Carena:2005gq}, without loss of generality. Expanding about the background metric, $G_{MN}= G_{MN}^0+h_{MN}$, with zeroth order metric $G_{\mu\nu}^0=e^{-2ky}\eta_{\mu\nu}$, and $G_{55}^0=1$ with
$G_{\mu5}^0=h_{\mu5}=0$ in a straight gauge, the boundary conditions give (indices are raised with $g^{\mu\nu}=e^{2ky}\eta^{\mu\nu}$):
\bea
& &\left[\frac{v_i}{2k}\left\{h_{\alpha\mu,\nu}^{\ \ \ \ \alpha}
    +h_{\alpha\nu,\mu}^{\ \ \ \ \alpha}-h_{\mu\nu,\alpha}^{\ \ \ \ \alpha}-\tilde{h}_{,\mu\nu}-g_{\mu\nu}(h_{\alpha\beta,}^{\ \ \
      \ \alpha\beta}-\tilde{h}_{,\alpha}^{\ \alpha})\right\}\right.\nonumber\\
& &\qquad \qquad+\ \left. \theta_i\left\{2kh_{\mu\nu}+h_{\mu\nu,5}-g_{\mu\nu}\tilde{h}_{,5}-3kg_{\mu\nu}h_{55}\right\}\frac{}{}\right]_{y=y_i}=0.
\eea

For massive 4D modes the tensor $h_{\mu\nu}$ can be written as
\bea
h_{\mu\nu}\rightarrow h_{\mu\nu}+\partial_\mu V_\nu +\partial_\nu
V_\mu
+e^{-2ky}\partial_\mu\partial_\nu\mathcal{S}_1+G^0_{\mu\nu}\mathcal{S}_2,
\eea
where $h_{\mu\nu}$ is now transverse-traceless with five degrees of freedom, $\partial^\alpha
h_{\alpha\beta}=\eta^{\alpha\beta}h_{\alpha\beta}=0$, and $V_\mu$ is
transverse, $\partial^\alpha V_\alpha=0$. Also $\mathcal{S}_1$ and
$\mathcal{S}_2$ are scalar degrees of freedom. One can show that the
physical massive modes are contained in $h_{\mu\nu}$~\cite{George:2011sw}. Performing a gauge transformation, with 4D gauge parameter $\xi_\mu$, the transverse component of $\xi_\mu$ is used to gauge away $V_\mu$, while the longitudinal part removes one of the scalars. The boundary conditions force the remaining scalar to vanish, absent fine-tuning among the brane curvature terms~\cite{George:2011sw}.

Writing the KK expansion for the physical fluctuations as
\bea
h_{\mu\nu}(x,y)=\kappa_*\sum_n h^{(n)}_{\mu\nu}(x)f_n(y),
\eea
where $\kappa_*$ is chosen to give the 4D fields
$h_{\mu\nu}^{(n)}$ a canonical mass dimension, the solution in the bulk  is
\bea
f_n(y)=\frac{1}{N_n}\left\{J_2\left(\frac{m_n}{k}e^{ky}\right)
+\beta_n Y_2\left(\frac{m_n}{k}e^{ky}\right)\right\},
\eea
with $m_n$ the mass of the $n^\text{th}$ spin-2 KK mode.
Applying the boundary conditions gives
\bea
\beta_n^i=-\frac{ J_1(z_i)
  -(z_iv_i\theta_i/2)J_2(z_i)}{ Y_1(z_i)
  -(z_iv_i\theta_i/2)Y_2(z_i)},\label{massive_KK_beta}
\eea
where $z_i=m_ne^{ky_i}/k$. The KK masses follow by enforcing $\beta_n^\uv=\beta_n^\ir\equiv \beta_n$. The mass for light IR-localized KK modes has a negligible dependence on the UV brane term - one can essentially take $v_\uv\approx0$ without modifying the spectrum. On the other hand, the IR term $v_\ir$ modifies the KK masses in a non-trivial way~\cite{Davoudiasl:2003zt}. For $v_i\rightarrow0$ the KK masses reduce to the usual RS values~\cite{Davoudiasl:1999jd}. We note that the additional factors of $1/2$ in Eq.~\eqref{massive_KK_beta}, relative to Ref.~\cite{Davoudiasl:2003zt}, can be removed by rescaling the value of $v_i$ in Eq.~\eqref{gravity_action}. This factor reflects the use of an interval rather than an orbifold (much as the brane tensions in Eq.~\eqref{gravity_action} are  smaller by a factor of $1/2$, relative to the orbifold picture). This scaling would introduce a factor of 2 in many equations below, so it is simpler not to rescale.  In our notation, the limit of large IR curvature gives a lightest KK graviton with mass approximately given by $m_{G_1}\approx 2e^{-kL}k/\sqrt{v_\ir/2}$. We note that the IR curvature of $r_L=10/k$ in Ref.~\cite{Falkowski:2016glr} corresponds to $v_\ir=20$, while $r_L=7$ in Ref.~\cite{Dillon:2016fgw} corresponds to $v_\ir\approx14$, and values of $\gamma_\pi\approx-7.6<0$ in Ref.~\cite{Hewett:2016omf} correspond to $v_\ir>0$, due to a notational difference.

The spin-2 spectrum contains the usual UV-localized massless graviton, with profile:
\bea
f_0(y)=e^{-2ky}\sqrt{ \frac{2k}{ 1-e^{-2kL}+\sum_i v_ie^{-2ky_i}}},\label{eq:graviton_profile}
\eea
A further massless mode (the scalar
radion) is present in the spectrum. This state acquires mass once the length
of the extra dimension is stabilized, with the corresponding mass dependent on the
backreaction of the stabilizing dynamics~\cite{Csaki:2000zn},  as we discuss below for
the Goldberger-Wise mechanism. However, it shall prove instructive
to first comment on the massless radion, as some results remain useful in the weak backreaction case. 

Thus, turning our attention  to the gravi-scalar fluctuations, we note that in a straight gauge one can always use remnant gauge freedom to write
the metric fluctuation $h_{55}$ as~\cite{Carena:2005gq,Bao:2005bv}
\bea
h_{55}(x^\mu,y)=F(y) \psi(x^\mu).\label{h_55_fluctuation}
\eea
Here $F(y)$ is an arbitrary function of $y$ satisfying $\int_0^Ldy\
F(y)\ne0$. An arbitrary $h_{55}$ can be cast into the form
(\ref{h_55_fluctuation}) via a general 5D coordinate
transformation, $x^M\rightarrow x^M+\xi^M$, with $\xi^\mu=0$ and~\cite{Carena:2005gq,Bao:2005bv}
\bea
\xi^5=\frac{1}{2}\int^ydy\, h_{55} -\frac{1}{2}\int^ydy\, F(y) \psi.
\eea
The presence of the arbitrary function $F(y)$ is a remnant gauge freedom.

We find it convenient to
write the most general form of the metric with background and scalar perturbations as
\bea
G_{MN}=\begin{pmatrix}
    a^2\left[\eta_{\mu\nu}
        +\nabla_\mu\nabla_\nu P_3
        +\eta_{\mu\nu}\left(2P_2-aa'P_3'\right)\right] & 0 \\
    0 & 1+2P_1-\left(a^2P_3'\right)'
\end{pmatrix} \:,
\label{p_metric_ansatz}
\eea
Here $a(y)$ is the  background warp factor and $P_{1,2,3}$ are spin-zero perturbations that are functions of $x^\mu$ and $y$.  This parametrization
is motivated by the gauge-invariant forms of Ref.~\cite{Bridgman:2001mc,Deffayet:2002fn}, and is such that  the Einstein
equations have a simple structure. For a detailed discussion of the gauge freedoms and the gauge transformations  that allow one to write the scalar perturbations in this form, see the Appendix in Ref.~\cite{George:2011sw}. Two of
the bulk Einstein equations can be cast as
\bea
\partial_\mu\partial_\nu\left(P_1+2P_2\right) &=& 0 \qquad  \mu\ne\nu \:,\label{einstein_relate_p1p2}\\
\partial_\mu\left(\frac{a'}{a}P_1 - P_2'\right) &=& 0 \qquad  \forall\mu\:.\label{einstein_P1_y-dep}
\eea
Taking the integration constants to vanish (the perturbations are localized in $x$), Eq.~\eqref{einstein_relate_p1p2}
relates $P_2$ and $P_1$, while Eq.~\eqref{einstein_P1_y-dep} determines the $y$-dependence of $P_1$.  The remaining bulk Einstein
equation reduces to $\Box P_1=0$, as expected for a massless 4D field.
The perturbation $P_3$ is completely free in the bulk, reflecting the
remnant gauge freedom~\cite{George:2011sw}. This is a
related to the remnant gauge freedom in the massless sector described
in~\cite{Carena:2005gq,Bao:2005bv},
and physical quantities do not depend on (the bulk value of) $P_3$. Boundary
conditions derive from the two additional
boundary Einstein equations:
\bea
P_3'(y_i)=\frac{-v_i}{a(y_i)\left[\theta_ika(y_i)+v_ia'(y_i)\right]} P_1(y_i) \:.
\label{p1p3_boundary_condition}
\eea

Using the solutions to the above one can compute the effective 4D
action for the physical scalar fluctuation.  We
perform separation of variables and solve for the profile of $P_1$,
giving
\bea
P_1= a^{-2}(y)\psi(x^\mu) \: \label{p1_expression}.
\eea
This solution is consistent with the boundary
conditions~\eqref{p1p3_boundary_condition} for general $v_i$
provided  the arbitrary function $P_3$ has  $P_3'\ne 0$ at the boundaries, in accordance with Eq.~\eqref{p1p3_boundary_condition}.
For the sources in Eq.~(\ref{gravity_action}), the solution for the background
metric has the standard RS form, $a(y)=e^{-ky}$.
Ignoring 4D surface terms ($\psi$ vanishes at
$x^\mu\rightarrow\infty$), and inserting the solution into the action, one obtains the effective 4D action for scalar
perturbations, up to $\mathcal{O}(\psi^2)$, as
\bea
\mathcal{S}_{\mathcal{O}(\psi^2)} = \int d^4x \left[
    \frac{3M_*^3}{k} e^{2kL}
    \left(\frac{1}{1-v_\ir} - \frac{e^{-2kL}}{1+v_\uv}\right)
  \right]
  \left( -\frac12 \eta^{\mu\nu} \partial_\mu \psi \partial_\nu \psi \right) \:\label{eq:radion_kinetic_GI}.
\eea
Note that linear terms in
$\psi$ and additional quadratic terms, which appear at intermediate stages of the calculation, cancel out in the final result, providing a check. In particular, 
higher-order derivative terms present at the quadratic level in
individual terms in Eq.~(\ref{gravity_action}), cancel out in
the full action.

The physical radion is defined as $r(x)=\psi(x) N_\psi$, where
the normalization constant is:
\bea
N_\psi^2=\frac{k}{3M_*^3}e^{-2kL}\frac{(1-v_\ir)(1+v_\uv)}{(1+v_\uv)-(1-v_\ir)e^{-2kL}}= \frac{e^{-2kL}}{6 M_{Pl}^2}(1-v_\ir)(1+v_\uv).\label{eq:unstable_Normalization}
\eea
We immediately observe from Eq.~\eqref{eq:radion_kinetic_GI} that the kinetic term is only well behaved for $v_\ir<1$, while the UV term suffers
no such constraint  (one may safely take $v_\uv\gg1$, as in Ref.~\cite{George:2011sw}). We assume that a ghost-like radion (a
wrong sign kinetic term) signals an instability of the ground state of
the theory and that it is desirable to fix this in the traditional way
by adding terms to the theory and/or restricting the couplings. Note that the crossover region between parameter space with/without a ghost-radion gives a vanishing kinetic term, meaning the theory is strongly coupled; such regions should also be avoided. Regarding parameter space with $(1+v_\uv)<0$, one should use Eq.~\eqref{eq:radion_kinetic_GI}, rather than Eq.~\eqref{eq:unstable_Normalization}, to determine whether problems arise,  due to the $v_\uv$-dependence of $M_{PL}$. Observe that the radion kinetic term is not problematic for $(1+v_\uv)<0$, whereas it is sensitive to the IR term, opposite to the massless graviton (this has a clear interpretation in the dual 4D theory, as discussed below).

It is worth emphasizing a point made above, with regard to the fluctuation $P_3$.  In the limit $v_i\rightarrow0$, one can use the
remaining gauge freedom to choose the form of the scalar fluctuations
such that the derivative pieces in Eq.~\eqref{p_metric_ansatz} vanish, namely
$\nabla_\mu\nabla_\nu P_3$ and $P_3'$~\cite{George:2011sw}.
Thus, in the limit of vanishing brane curvature, the standard
parametrization of the gravi-scalar metric fluctuations in RS~\cite{Goldberger:1999un,Csaki:2000zn,Cheung:2000rw} is found to be  consistent
with the boundary conditions in the interval picture. However, for $v_i\ne0$, one
is unable to remove the derivative pieces in Eq.~\eqref{p_metric_ansatz} with a
gauge choice while \emph{simultaneously} obtaining a solution that is consistent with the
boundary conditions~\cite{George:2011sw}. 
%%%%%%%%%%%%%%%%%%%%%%%%%%%%%%%%%%%%%%%%%%%%%%%%%%%%%%%%%%%%%%55

%%%%%%%%%%%%%%%%%%%%%%%%%%%%%%%%%%%%%%%%%%%%%%%%%%%%%%%%%%%%%%55
%%%%%%%%%%%%%%%%%%%%%%%%%%%%%%%%%%%%%%%%%%%%%%%%%%%%%%%%%%%%%%%%%%%5
\section{ Radion Coupling to Brane Matter\label{sec:radion-matter-GI}}

We now turn to the coupling of the radion to brane-localized matter, which depends on the location of the matter. Some expressions presented below  generalize results of Ref.~\cite{George:2011sw} for the case with IR curvature. Consider a set of matter fields localized at the boundary $y=y_i$.  Expanding the metric in terms of a fluctuation $f_{\mu\nu}$, which
only contains  the spin-zero parts of the perturbation, 
$g_{\mu\nu}\rightarrow g_{\mu\nu}+f_{\mu\nu}$, integrating over
the extra dimension,  and scaling the matter fields to bring the kinetic terms to canonical form, the linear fluctuation term is
\bea
\left.S\right|_{\mathcal{O}(f)}=-\frac{1}{2}e^{2k y_i} \int
d^4x\ 
\eta^{\mu\alpha}\eta^{\nu\beta}f_{\mu\nu}\ T_{\alpha\beta} \:, \label{fluc_coupling_T}
\eea
where $T_{\mu\nu}$ is written in terms of the flat space metric (and canonical fields). Consider the non-derivative couplings of the gravi-scalar  to $T_{\mu\nu}$:
\bea
\left.S\right|_{\mathcal{O}(\psi)}=\frac{e^{2ky_i}}{2}\left[1-\frac{v_ia'}{(\theta_ika+v_ia')}\right]\int
d^4x\ \psi\ T+\dots \label{S_f_general gauge_GI}.
\eea
 where $T=\eta^{\mu\nu}T_{\mu\nu}$ and we used
Eqs.~(\ref{einstein_relate_p1p2})-(\ref{p1_expression}). The coupling of the physical radion $r$ 
 is 
\bea
\left.S\right|_{\mathcal{O}(r)}=\frac{1}{2} \int
d^4x\ \left(\frac{r}{\Lambda_i}\right)\times T+\dots\ ,\label{S_f_L gauge_GI}
\eea
with location-dependent coupling $\Lambda$. For matter localized on the brane at $y=y_i$, one has
\bea
\Lambda_{i}^{-1}&=&e^{kL}\left\{{\frac{k}{3M_*^3}}\frac{1}{(1+v_\uv)-(1-v_\ir)e^{-2kL}}\right\}^{1/2} \sqrt{\frac{1-\theta_jv_j}{1-\theta_i v_i}}\nonumber\\
&=& \frac{1}{\sqrt{6}}\; \frac{1}{e^{-kL} M_{Pl}}\times  \sqrt{\frac{1-\theta_jv_j}{1-\theta_iv_i}}\quad\quad i\ne j.\label{eq:IR_matter_r_coupling}
\eea
One can summarize the brane radion couplings as
\bea
\Lambda_{i}&=&\Lambda_{RS,i}\times  \sqrt{\frac{1-\theta_i v_i}{1-\theta_jv_j}}\quad\quad i\ne j.
\eea
where $\Lambda_{RS,i}$ is the usual RS radion coupling for matter on the brane at $y_i$. Thus, in the limit $v_{\uv,\ir}\rightarrow0$ one obtains the standard RS results. Note that the IR coupling is
\bea
\Lambda_{\ir}^{-1}&=&\Lambda_{RS,\ir}^{-1}\times  \sqrt{\frac{1+ v_\uv}{1-v_\ir}},
\eea
which becomes strongly coupled for $v_\ir\rightarrow1$. This corresponds to the crossover region between having/avoiding a ghost-like radion, such that the kinetic term vanishes, as mentioned previously.

At first sight the $v_\ir$ dependence of these couplings appears unusual. Intuitively, one may expect the IR coupling to diminish for increasing values of $v_\ir$, and the UV coupling to have limited sensitivity to the size of $v_\ir$. However, one observes that increasing values of $v_\ir$ tend to decrease the coupling at the UV brane and increase the coupling at the IR brane. Actually, this behaviour is not so surprising. Recall that increasing values of $v_\ir$ cause  the strength of the  kinetic term for the unscaled fluctuation $\psi = r/N_\psi$ to increase; see Eq.~\eqref{eq:radion_kinetic_GI}. After scaling $\psi$, this translates into a suppression of the couplings to the radion $r$, for increasing $v_\ir$. For UV localized matter, this is the only $v_\ir$ dependence in the coupling, giving the inverse sensitivity of $\Lambda_{\uv}^{-1}$ to $v_\ir$. Note that for $v_\uv\rightarrow 0$, the IR coupling has a simple form, $\Lambda_\ir=\Lambda_{RS,\ir}\times\sqrt{1-v_\ir}$, while for $v_\ir\rightarrow1$ one has $\Lambda^{-1}_\ir\rightarrow\infty$, and the theory enters a strong coupling regime. We comment more on this coupling in Section~\ref{sec:ads/cft}.

As an additional point, we note that Eq.~\eqref{eq:IR_matter_r_coupling} displays the expected dependence on the UV curvature  $v_\uv$. In the limit $v_\uv\rightarrow\infty$, the UV coupling vanishes, $\Lambda^{-1}_\uv\rightarrow0$,  with the radion  expelled from the UV brane. This corresponds to decoupling 4D gravity by sending the 4D Planck scale to infinity. On the other hand, the limit $v_\uv\rightarrow\infty$ has little effect on the IR coupling, which remains as $\Lambda_\ir\sim e^{-kL} M_*$, namely the characteristic IR scale. This makes intuitive sense, given the interpretation of the radion as a composite dilaton in the dual picture.

The
preceding discussion is relevant for brane localized matter. However, it retains utility for models  with bulk fields. The radion couples
conformally to matter. In models where SM
fermions are treated as zero modes of bulk fermions, they
 typically remain massless until electroweak symmetry breaking is triggered by an IR localized Higgs. Consequently fermion masses arise locally on the IR brane. The mass-induced coupling between the radion and  SM
fermions therefore occurs locally on the IR brane, with a strength
controlled by Eq.~\eqref{eq:IR_matter_r_coupling}, giving $\Lambda^{-1}_\ir\sim e^{kL}\sqrt{k/M_*^3}$. This statement is not sensitive to the localization profile of the
zero-mode fermion; information regarding the 
wavefunction overlap
 with the IR brane is encoded in the effective
4D fermion mass. The radion coupling to a fermion $f$ goes
like 
$(m_f/\Lambda_r)\times r\bar{f}f$, being smaller for an electron than a top
quark simply because $m_e\ll m_t$, regardless of the origin of this hierarchy (e.g., tiny input Yukawa couplings or suppressed wavefunction overlap). Similar discussion holds for zero modes of bulk vectors that acquire mass from an IR localized scalar.

%%%%%%%%%%%%%%%%%%%%%%%%%%%%%%%%%%%%%%%%%%%%%%%%%%%%%%%%%%%%%%%%%%%5
\section{The Radion in GW Stabilized RS Models\label{sec:GW}}
In the preceding sections the radion was massless as no mechanism was employed to stabilize the length of the extra dimension.
Here we briefly discuss the case where the radion acquires mass due to radius stabilization via the Goldberger-Wise mechanism~\cite{Goldberger:1999uk}. This approach introduces a bulk scalar $\Phi$, with localized boundary potentials, to generate a potential for the length of the interval. The result is a KK tower of physical scalars that contain an admixture of the KK modes of $\Phi$ and the gravi-scalar. The radion is identified as the lightest mode in this KK tower. 

With GW scalar included, the complete action is 
\begin{align}
\mathcal{S} &=
  \int_{\mathcal{M}} d^5x \sqrt{-G}
    \left\{2M_*^3 \mathcal{R}
      - \frac12 G^{MN}\partial_M\Phi\partial_N\Phi - V(\Phi)\right\} + 4M_*^3 \oint_{\partial \mathcal{M}} \sqrt{-g}\, K \nonumber\\
&\qquad
  + \sum_i \int d^4x \sqrt{-g_i}
    \left\{\frac{M_*^3v_i}{k} R_i - M_*^3k w_i
      - \frac14 t_i g^{\mu\nu} \partial_\mu\Phi\partial_\nu\Phi - \frac12 \lambda_i(\Phi)\right\}. 
\label{5d_stabilized_action}
\end{align}
We include brane kinetic terms for both the gravity
($v_i$) and scalar ($t_i$) sectors.  $V(\Phi)$ is the bulk potential for the
scalar $\Phi$ (which subsumes the bulk cosmological constant) and
$\lambda_i$ are brane localized potentials.
The brane tensions $kw_i$ are explicitly separated from the brane
potentials, so $\lambda_i(\Phi)=0$ for the background value of $\Phi$. The general analysis of this system was presented in Ref.~\cite{George:2011sw}. Here, we summarize a few key results, which we subsequently generalize. For detailed discussion of the methodology see Ref.~\cite{George:2011sw}.

Taking the usual warped metric ansatz:
\bea
ds^2=a^2(y)\eta_{\mu\nu}dx^\mu dx^\nu +dy^2,
\eea
where the warp factor $a(y)$ is to be determined, and allowing the background value
of $\Phi$ to depend
only on $y$,\footnote{That is, we write $\Phi(x^\mu,y)=\phi(y) +P_4(x^\mu,y)$, where  $\phi(y)$ is the background value for $\Phi$. See Ref.~\cite{George:2011sw} for more details.}, one can obtain the equations of motion and boundary conditions for the combined gravity-scalar theory. Two of the boundary conditions remain as in Eq.~\eqref{p1p3_boundary_condition}, while the other two have the form 
\bea
\left[
  t_i\partial_\mu(\sqrt{-g}\,g^{\mu\nu}\partial_\nu\Phi)
    - \sqrt{-g}\,\lambda_{i,\Phi}
    - 2\theta_i\sqrt{-G}\,G^{5N}\partial_N\Phi
\right]_{y=y_i} = 0 \:,
\eea
when expressed in straight gauge.\footnote{Additional useful forms of the boundary conditions appear in Ref.~\cite{George:2011sw}.} The effective 4D theory contains the following terms for the KK scalars:
\bea
\mathcal{S} &\supset&\mathcal{N} \int d^4x \, \left(
  -\frac12 \eta^{\mu\nu}\partial_\mu\psi\partial_\nu\psi
  -\frac12 m^2 \psi^2 \right)  \:,
\eea
where $m^2$ is the mass of the KK mode and the normalization factor is
\begin{align}
\mathcal{N} &=
  6M_*^3 \int_0^L \left(
    a^2 p_1^2
    + 24M_*^3 \,\frac{a'^2}{\phi'^2} p_1^2
    + 24M_*^3\,\frac{aa'}{\phi'^2} p_1 p_1'
    + 6M_*^3\,\frac{a^2}{\phi'^2} p_1'^2
    \right) dy\nonumber\\&\qquad
  + 3M_*^3 \sum_i \frac{v_i a(y_i)^3 p_1(y_i)^2}{ka(y_i)+\theta_iv_ia'(y_i)}
\nonumber\\&\qquad
  + \frac18 \sum_i t_i\left[
    12M_*^3\,\theta_i\frac{2a'(y_i)p_1(y_i)+a(y_i)p_1'(y_i)}{\phi'(y_i)}
    +\frac{v_ia(y_i)^2\phi'(y_i)p_1(y_i)}{ka(y_i)+\theta_iv_ia'(y_i)}\right]^2 \:.
\label{stabilised_radion_norm}
\end{align}

In the above, the
 form of the background metric is not specified. The point is that the potentials
 $V(\Phi)$ and $\lambda_i(\Phi)$ cause $\Phi$ to obtain a nontrivial
 background value, which combines with the bulk cosmological constant and
 the brane tensions to source the metric.  To calculate the radion mass, one must specify a particular model by specifying the  form for the background scalar. To allow comparison with existing results in the literature, we follow Ref.~\cite{Csaki:2000zn} and consider a perturbed background of the form
\bea
a(y) &=& e^{-ky}\left(1-\frac{l^2}{6}e^{-2uy}\right) ,\\
\phi(y) &=& 2\sqrt{2}M_*^{3/2}\,le^{-uy} \:,
\eea
valid in the region $y\in[0,L]$. This corresponds to a
potential $V(\Phi)=(W_{,\Phi})^2/2-W^2/6M_*^3$
with $W(\Phi)=12M_*^3k-u\Phi^2/2$, and the following boundary potentials
\bea
\lambda_i(\Phi)=-\theta_i\,W(\phi_i)-\theta_i\,W_{,\Phi}(\phi_i)(\Phi-\phi_i)+\gamma_i(\Phi-\phi_i)^2,
\eea
with constants $u$, $\phi_i$ and $\gamma_i$. The length of the extra
dimension is now dynamically  fixed at $L=u^{-1}\log(\phi_0/\phi_L)$, with the weak backreaction
limit defined by $\kappa_*\phi_i/\sqrt{2}\ll1$. We work to $\mathcal{O}(l^2)$ in the small parameter $l=\kappa_*\phi_0/\sqrt{2}$, though the expression for $\phi$ holds to all
  orders in $l$. 

Writing the metric perturbation as $P_1(x^\mu,y)=p_1(y) \psi(x^mu)$, the solution for $p_1(y)$ is a perturbed
form of the massless solution,
\bea
p_1(y) &=& \left\{1+l^2f(y)\right\} \times e^{2ky}.
\eea
The bulk equation for $f(y)$ is the same as the case without brane curvature terms~\cite{Csaki:2000zn},
\bea
f'' + 2(k+u)f' &=& \frac43u(u-k)e^{-2uy} - \tilde{m}^2e^{2ky} ,
\eea
where $m^2=l^2\tilde{m}^2$. Observe that the radion mass is on the order of the
correction to the background -  the backreaction must be included to generate a non-zero mass. The solution in the bulk is~\cite{Csaki:2000zn}
\bea
f'(y) &=& -\frac23u\left(1-\frac{u}{k}\right)e^{-2uy}
  - \tilde{m}^2\frac{1}{4k+2u}e^{2ky}
  + Ae^{-2(k+u)y} ,
\label{radion_solution}
\eea
where $A$ is an integration constant. Working in the limit of stiff brane potentials, $\lambda_{i,\Phi\Phi}\to\infty$, the boundary conditions are~\cite{George:2011sw}
\bea
\left[ f' + \frac23ue^{-2uy}
  + \frac23\frac{u^2}{k}e^{-2uy} \frac{\theta_iv_i}{1-\theta_iv_i}
\right]_{y=y_i} &=& 0 ,\label{eq:radion_f_BC}
\eea
enforcing which allows one to determine 
the mass of the lightest spin-zero state~\cite{George:2011sw}
\bea
m^2&=&\frac{4l^2}{3}\frac{(2k+u)u^2}{k}
  \left(\frac{1}{1-v_\ir} - \frac{e^{-2kL}}{1+v_\uv}\right)
  \left(e^{2(k+u)L}-e^{-2kL}\right)^{-1} \label{radion_mass}
\eea
This expression for the radion mass generalizes of the result in
Ref.~\cite{Csaki:2000zn} for the case of
non-zero brane curvature, $v_i\ne0$. There are two points worth making. Firstly, one observes that $v_\ir<1$ is required for the theory to remain consistent. In particular, values of $0<v_\ir<1$ tend to increase the mass of the radion,
relative to the standard RS result. This differs from the case of the KK gravitons, where the increase in $v_\ir$ corresponds to a reduction in the KK masses (as used recently in relation to the 750~GeV diphoton excess). Secondly, while the mass is sensitive to the effect of the IR curvature, $v_\ir$, it is rather insensitive to the UV curvature $v_\uv$.  

 In cases where it is desirable to have a heavy radion, Eq.~\eqref{radion_mass} might lead one to suppose that we could use the IR  curvature  to achieve this while avoiding large perturbations to the AdS background.  However, one must be careful, as although values of $v_\ir$ close to unity enhance the radion mass, they also approach a strongly coupled regime in the radion interactions (corresponding to the crossover region between having/avoiding a ghost-like radion, where the kinetic term vanishes).

To determine the coupling of the radion to matter in the stabilized extra dimension, one requires the normalization constant $\mathcal{N}$ in
Eq.~\eqref{stabilised_radion_norm}. Unsurprisingly, we find 
\bea
\mathcal{N} =
  \frac{3M_*^3}{k}e^{2kL}
    \left(\frac{1}{1-v_\ir}-\frac{e^{-2kL}}{1+v_\uv}\right)
    + \mathcal{O}(l^2) \:,
\label{massive_radion_norm_leading_order}
\eea
matching the non-stabilized result in
Eq.~\eqref{eq:unstable_Normalization} to leading order. With this expression one can repeat the
calculations of Section~\ref{sec:radion-matter-GI} to find the
radion coupling to matter. To leading order the results match those in Section~\ref{sec:radion-matter-GI}.

%%%%%%%%%%%%%%%%%%%%%%%%%%%%%%%%%%%%%%%%%%%%%%%%%%%%%%%%
\section{Radion-Higgs Kinetic Mixing\label{sec:radion_higgs_mixing}}
In the above, we considered a GW stabilized RS model with brane curvature. The scalar spectrum consisted of a KK tower of massive scalars, the lightest of which is the radion. For model building purposes one would subsequently add the SM fields to the warped space. In particular, one adds the SM Higgs boson, which should be localized at (or towards) the IR brane in order to solve the hierarchy problem. In general the Higgs will mix with the KK scalars, the most important consequence of which is the mixing between the Higgs and the lightest mode, namely the radion. In addition to the evident phenomenological implications, the radion-Higgs mixing has a further consequence. We observed previously that the radion mass in the GW stabilized setup is only well behaved for relatively moderate values of the IR curvature, $v_\ir<1$. This observation is important with regards to efforts to suppress the lightest KK graviton mass by employing values of $v_\ir\sim\mathcal{O}(10)$, as such models could suffer from instabilities. It is also important for models seeking to generate a 750~GeV radion, as the IR curvature term can be used to increase the radion mass. These conclusions, however, are drawn prior to the inclusion of the Higgs-radion mixing. In this section we discuss the modifications to these observations due to Higgs-radion mixing. The results in this section generalize a number of the results in Ref.~\cite{Csaki:2000zn}  to include non-zero brane curvature.

We are interested in the case where the SM is added to the warped space. For present purposes, we assume an IR localized SM Higgs:
\bea
S&\supset& -\int d^4x \sqrt{g_\ir} \{(D^\mu H)^\dagger (D_\mu H) + V(H)\}+ \xi \int d^4x \sqrt{g_\ir}\, |H|^2 R_\ir,
\eea
where we include an IR-localized non-minimal coupling. There are two ways one can proceed to analyze the system of Section~\ref{sec:GW} with the SM Higgs added. The most-general analysis involves deriving the full equations of motion and boundary conditions for the gravity+$\Phi$+$H$ system, and deriving the new KK spectrum for the scalar sector (comprised of the radion, $\Phi$ and $H$). Alternatively one can treat the SM Higgs as a small perturbation on the previously derived background solution and derive the leading order mixing effects between the Higgs and the KK scalars. Here we make a simple observation which allows an intermediate approach.

Expanding the Higgs around its VEV, one has 
\bea
S&\supset&\frac{\xi}{2} \int d^4x \sqrt{g_\ir}\, e^{2kL}\,(h^2 +2 v h +v^2)\, R_\ir,\label{eq:IR_nonminimalcoupling}
\eea
where the Higgs is rescaled to the canonical kinetic form, $H\rightarrow e^{kL}\,H$, with $v\ll M_{Pl}$ being the warped-down SM VEV, $v\simeq 246$~GeV. Observe that the non-minimal IR coupling gives two different physical effects. The $\mathcal{O}(v)$ term induces Higgs-radion kinetic mixing, requiring one to diagonalize the scalar kinetic terms, as discussed below. On the other hand, the $\mathcal{O}(v^2)$ term does not induce kinetic mixing but instead gives a new contribution to the \emph{total} IR brane curvature. To treat this term, we can rewrite the IR curvature as
\bea
S&\supset& \frac{M_*^3}{k}  \int d^4x \sqrt{-g_\ir}
    \left\{v_\ir+\xi v_H\right\}R_\ir, 
\eea
with the dimensionless coefficient $v_H= (k v^2)/(2e^{-2kL} M_*^3)$ parametrizing the Higgs contribution to the effective IR-localized curvature. This makes it clear that the $\mathcal{O}(v^2)$ term in the non-minimal coupling can be incorporated in our earlier analysis by the replacement $v_\ir\rightarrow v_\ir +\xi v_H$ in the action \eqref{5d_stabilized_action}. The results obtained via this approach reduce to those obtained by the alternative method of treating this term as a perturbation (see the Appendix). Let us also emphasize that the KK graviton masses are sensitive to the total IR curvature, and are thus sensitive to the value of $v_H$ for $\xi\ne0$.

With the above observation one easily includes the effects of the Higgs-induced IR curvature into the full equations of motion and boundary conditions, following the analysis of Ref.~\cite{George:2011sw} (as outlined in the preceding section). To quote a few key results, the expression for the Planck mass becomes 
\bea
M_{Pl}^2=\frac{M_*^3}{2k}\left\{1+v_\uv - (1-v_\ir-\xi v_H)e^{-2kL} \right\},\label{eq:4D_planck_nonminimal}
\eea
along with a related  change to the massless graviton profile. One of the IR boundary conditions changes to
\bea
P_3'(y_\ir)=\frac{-(v_\ir +\xi v_H)}{a(y_\ir)\left[ka(y_\ir)+(v_\ir +\xi v_H) a'(y_\ir)\right]} P_1(y_\ir) \:,
\label{p1p3_boundary_condition_new}
\eea
and the leading-order expression for the radion normalization factor becomes:
\bea
\mathcal{N} =
  \frac{3M_*^3}{k}e^{2kL}
    \left(\frac{1}{1-(v_\ir +\xi v_H)}-\frac{e^{-2kL}}{1+v_\uv}\right)
    + \mathcal{O}(l^2) \:,
\label{nonminimal_radion_norm_leading_order}
\eea
The stiff brane-potential  limit expression for the boundary conditions, Eq.~\eqref{eq:radion_f_BC},  also changes, and the new $\mathcal{O}(l^2)$ expression for the radion mass is
\bea
m^2&=&\frac{4l^2}{3}\frac{(2k+u)u^2}{k}
  \left(\frac{1}{1-v_\ir -\xi v_H} - \frac{e^{-2kL}}{1+v_\uv}\right)
  \left(e^{2(k+u)L}-e^{-2kL}\right)^{-1} \:.\label{nonminimal_radion_mass}
\eea

With the $\mathcal{O}(v^2)$ term incorporated into  the full equations of motion and boundary conditions, we now treat the $\mathcal{O}(v)$ term as a perturbation on the new background solution.\footnote{Note that we are not performing an expansion in the  parameter $v$ here; references to  $\mathcal{}O(v)$ and $\mathcal{O}(v^2)$ terms in the non-minimal coupling are made purely for labeling purposes.} The calculation makes use of the following result for the linear terms in the radion fluctuation:
\bea
\sqrt{g_i} R_i&=& \frac{3k a^3(y_i) p_1(y_i)\Box \psi}{ka(y_i)+\theta_i v_i a'(y_i)} +\ldots,
\eea
where one should use the total brane curvature for the IR brane, $v_i\rightarrow v_\ir+\xi v_H$.
Using this result to extract the kinetic mixing gives
\bea
S&\supset&\frac{3\xi}{(1-v_\ir-\xi v_H)}\int d^4x e^{2kL} v\,h\, \Box \psi. \label{eq:radion_higgs_kinetic_mixing}
\eea
We perform a partial rescaling of the radion kinetic term,
\bea
\psi\ =\  r\times\sqrt{\frac{(1+v_\uv)}{6 M_{Pl}^2e^{-2kL}}},
\eea
such that the kinetic mixing term is
\bea
S&\supset&3\xi\, \frac{v}{\Lambda_{RS,\ir}} \frac{\sqrt{1+v_\uv}}{(1-v_\ir-\xi v_H)}\; \int d^4x \,h\, \Box r \ \equiv\ \frac{A}{B}\int d^4x \,h\, \Box r,
\eea
where we define $B\equiv (1-v_\ir -\xi v_H)$. This partial scaling allows our results to be readily compared with Ref.~\cite{Csaki:2000zn}.

The mixed kinetic Lagrangian contains the terms
\bea
\mathcal{L}&\supset&\frac{1}{2}(r,\;h)
\left(
\begin{array}{cc}
B^{-1}&0\\
2AB^{-1}&1
\end{array}
\right)
\left(
\begin{array}{c}
\Box r\\
\Box h
\end{array}
\right),
\eea
which are diagonalized by the following $GL(2)$ transformation:
\bea
\left(
\begin{array}{c}
r\\
h
\end{array}
\right)
=
\left(
\begin{array}{cc}
\mathcal{Z}^{-1}&0\\
-A(B\mathcal{Z})^{-1}& 1
\end{array}
\right)
\left(
\begin{array}{c}
r'\\
h'
\end{array}
\right)
\eea
where 
\bea
\mathcal{Z}^2&\equiv&B^{-1}-(A/B)^{2}\nonumber\\
& =& \frac{1-v_\ir-\xi v_H- 9\xi^2\, \gamma^2 (1+v_\uv)}{(1-v_\ir-\xi v_H)^2}.\label{eq:Z2_radion_kinetic}
\eea
Here we adopt the notation of Ref.~\cite{Csaki:2000zn}: 
\bea
\gamma&=& \frac{v}{\Lambda_{RS,\ir}}\ =\ \frac{v}{\sqrt{6} e^{-kL} M_{Pl}}.
\eea
The quantity $\mathcal{Z}^2$ in Eq.~\eqref{eq:Z2_radion_kinetic} corresponds to the coefficient of the radion kinetic term after the kinetic-mixing is diagonalized. It should be strictly positive to ensure the kinetic term is positive definite and avoid a ghost-like radion.  Eq.~\eqref{eq:Z2_radion_kinetic} generalizes the result in Ref.~\cite{Csaki:2000zn}  for the case with localized brane curvature. We can consider various limits of this expression. The limit $\gamma^2\ll1$ gives
\bea
\mathcal{Z}^2& =& (1-v_\ir)^{-2} \left\{1-v_\ir+\frac{3\xi\, \gamma^2}{2}\left[1-6\xi (1+v_\uv)\right]\right\}+\ldots.\label{eq:Z2_radion_kinetic_smallgamma}
\eea
Taking the further limit of vanishing brane curvature, $v_{\ir,\uv}\rightarrow0$, gives
\bea
\mathcal{Z}^2& =&  1+\frac{3}{2}\xi\, \gamma^2(1-6\xi ),\label{eq:Z2_radion_kinetic_csaki_etal}
\eea
which matches the expression in Ref.~\cite{Csaki:2000zn}.\footnote{After correcting for a notational difference.} 

Returning the curvature terms, $v_{\ir,\uv}\ne0$, and demanding that the radion is not ghost-like, Eq.~\eqref{eq:Z2_radion_kinetic_smallgamma} shows that one should restrict $\xi$ to the range
\bea
\xi_-\ \le\ \xi\ \le\ \xi_+,
\eea
where
\bea
\xi_\pm&=& \frac{1}{12(1+v_\uv)}\left\{ 1\pm\left[1+\frac{16(1+v_\uv)(1-v_\ir)}{\gamma^2}\right]^{1/2}\right\}.
\eea
This expression also generalizes Ref.~\cite{Csaki:2000zn}. Note that it appears difficult to select values of $\xi$ consistent with $v_\ir\sim\mathcal{O}(10)$. The only hope arises for values of $0<(1+v_\uv)\ll1$, specifically, with $(1+v_\uv)=\epsilon \times\gamma^2/[16(v_\ir-1)]<1$ for small $\epsilon$.\footnote{Recall that $(1+v_\uv)$ must be strictly positive, to avoid a ghost-like massless 4D graviton.} However, this solution is misleading; it gives $\xi_\pm\propto (v_\ir-1) \gamma^{-2}$, such  that the original expansion in $\gamma\ll1$ cannot be trusted for $\xi$ in the range $\xi_-<\xi<\xi_+$, given that $\gamma$ is multiplied by a factor of $\xi$ or $\xi^2$ in Eq.~\eqref{eq:Z2_radion_kinetic_smallgamma}.  This failure to find values of $\xi$ that avoid a ghost-like radion for $v_\ir\sim\mathcal{O}(10)$ is best understood via Eq.~\eqref{eq:Z2_radion_kinetic}, which gives the more-general constraint for avoiding a ghost-like radion in the presence of brane curvature and a non-minimal coupling to an IR  Higgs, namely:
\bea
v_\ir+\xi v_H+ 9\xi^2\, \gamma^2 (1+v_\uv) \ < \ 1.\label{eq:no_ghost_radion_constraint}
\eea
One notes immediately that no solution with $\mathcal{O}(10)$ IR curvature appears possible, given that the last term on the left hand side is positive definite.  Naively one may expect that a cancellation could be arranged between the terms $v_\ir$ and $\xi v_H$, such that $v_\ir\sim\mathcal{O}(10)$ is allowed. However, the brane curvature relevant for modifying the KK graviton mass is the \emph{total} IR curvature, so a suppressed KK graviton mass requires $v_\ir+\xi v_H\sim\mathcal{O}(10)$ in the presence of the non-minimal coupling. Thus,  no such cancellation is available. We conclude that it appears difficult to reconcile the constraint in Eq.~\eqref{eq:no_ghost_radion_constraint} with $\mathcal{O}(10)$ values of the total IR brane curvature. Small values of the IR curvature, consistent with the above constraints, remain viable.

In the basis with Higgs-radion kinetic mixing, the scalar mass terms are diagonal: $\mathcal{L}\supset - m^2 r^2/2 - m_h^2 h^2/2$. However, diagonalizing the kinetic terms induces mass mixing. Defining the physical mass eigenstates as
\bea
\left(
\begin{array}{c}
r_m\\
h_m
\end{array}
\right)=
\left(
\begin{array}{cc}
\cos\theta&\sin\theta\\
-\sin\theta&\cos\theta
\end{array}
\right)
\left(
\begin{array}{c}
r'\\
h'
\end{array}
\right),
\eea
the mass eigenvalues are
\bea
m^2_\pm&=&\frac{1}{[2(B\mathcal{Z})^2]^{-1}}\left\{m^2 B^2 +m_h^2B\mathcal{Z} (A^2+B\mathcal{Z})\pm \Delta_m\right\}\nonumber
\eea
where
\bea
\Delta_m&=&  \left[(m^2 B^2 +m_h^2B\mathcal{Z} (A^2-B\mathcal{Z}))^2 -4m_h^4A^2(B\mathcal{Z})^2\right]^{1/2},
\eea
and, as previously, $m^2$ is the radion mass prior to mixing, $B=(1-v_\ir-\xi v_H)$, $A=3\xi\gamma \sqrt{1+v_\uv}$, and $\mathcal{Z}$ is defined  by Eq.~\eqref{eq:Z2_radion_kinetic}. The identification of the physical Higgs and radion with $m_\pm$ depends on the mass ordering; the physical radion has mass $m_+$ ($m_-$) if it is heavier (lighter) than the physical Higgs. The mixing angle is
\bea
\tan 2\theta &=& \frac{2AB\mathcal{Z}m_h^2}{m^2 B^2 +m_h^2B\mathcal{Z}(A^2-B\mathcal{Z})}.
\eea
Writing the above results for the mixing angle and eigenmasses in terms of the explicit expressions for $A$, $B$, and $\mathcal{Z}$ produces cumbersome expressions that are not particularly enlightening. One can consider various limits of the results, however. As an example, for $\gamma\ll1$, the mixing angle reduces to
\bea
\tan2\theta&=& \frac{6m_h^2 \xi \gamma\sqrt{1+v_\uv}}{m^2(1-v_\ir)-m_h^2},
\eea
which is the generalization of the result in Ref.~\cite{Csaki:2000zn} for $v_{\ir,\uv}\ne0$.

%%%%%%%%%%%%%%%%%%%%%%%%%%%%%%%%%%%%%%%%%%%%%%%%%%%%%%%%%%%%%5

%%%%%%%%%%%%%%%%%%%%%%%%%%%%%%%%%%%%%%%%%%%%%%%%%%%%%%%%
Our results show that the inclusion of Higgs-radion mixing via an IR localized non-minimal coupling does not provide a means for avoiding a ghost-like radion in a GW-stabilized RS model with large IR curvature. Of course, the radion mass is dependent on the back-reaction of the stabilizing dynamics and one may wonder if the ghost-like radion can be avoided in the case of a strong back-reaction, perhaps with different stabilizing dynamics. We have nothing insightful to say regarding this possibility, though we note that the strong back-reaction would also affect the mass of the KK gravitons. One may ask if the brane localized kinetic terms for the GW scalar can affect the sign of the radion kinetic term.  We note that the $t_i$-dependent terms in the normalization factors in Eq.~\eqref{stabilised_radion_norm}  give an $\mathcal{O}(l^2)$ correction to the leading-order radion normalization factor. Thus, in the weak backreaction limit one does not expect these terms to dominate the $\mathcal{O}(l^0)$ terms, and the ghost-like radion is expected to persist for $v_\ir\sim\mathcal{O}(10)$. Other possibilities include taking the Higgs ``off the wall"~\cite{Davoudiasl:2005uu} and into the bulk, or considering warped models with a different mechanism of stabilization. In Ref.~\cite{Cox:2013rva} the effects of bulk SM fields on the radion couplings were studied, however, it would be interesting to study this scenario with additional brane localised curvature terms.  Leaving these points aside, we now turn our attention to some alternative IR terms.
%%%%%%%%%%%%%%%%%%%%%%%%%%%%%%%%%%%%%%%%%%%%%%%%%%%%%%%%
\section{Additional IR Terms for the Radion\label{sec:new_ir_rad}}
In a certain sense, the use of branes in the RS model means the brane-localized action need only satisfy the 4D diffeomorphism symmetry. This allows a number of additional  terms that, in general,  should be included in the most-general Lagrangian. This fact was  already invoked to motivate the study of brane curvature terms and the non-minimal coupling to the IR Higgs. Motivated by the work of Ref.~\cite{Chacko:2003yp}, in this section we comment on a more speculative class of brane terms. In particular, we note that Ref.~\cite{Chacko:2003yp} considered explicit brane-localized mass terms for the spin-2 metric fluctuations $h_{\mu\nu}$.\footnote{Note that Ref.~\cite{Gherghetta:2005se} considered an explicit bulk mass for the graviton in RS models.} Such terms explicitly break the 5D general coordinate invariance and essentially force one to choose a gauge - the inclusion of such terms is clearly speculative. However, given recent interest in RS models with large IR curvature, and the inherent problem of the ghost-like radion encountered within, we feel it is prudent to comment on related speculations with regard to the radion. More precisely, we shall comment on localized terms for the scalar metric perturbation $h_{55}$, which preserve the local 4D symmetry but break the 5D general coordinate invariance. 

Consider the following additional terms on the IR brane:
\bea
\delta  \mathcal{S}&=&-3 \xi_\partial \frac{M_*^3}{k} \int d^4x \sqrt{-g_\ir} g^{\mu\nu} \partial_\mu h_{55}\partial_\nu h_{55} - \xi_m k \int d^4x \sqrt{-g_\ir} h_{55}^2 \Phi^2.\label{eq:new_ir_terms_rad}
\eea
Treating these terms as perturbations on the background, the first term is an IR localized kinetic term for $h_{55}$, which gives a new contribution to the kinetic Lagrangian:\footnote{Here $v_\ir$ can include the contribution from the IR Higgs, if desired.}
\bea
\delta  \mathcal{S}&\supset&-\frac{1}{2} \frac{8\xi_\partial}{(1-v_\ir)^2} \left(\frac{3M_*^3}{k} e^{2kL}\right) \int d^4x \,(\eta^{\mu\nu}\,\partial_\mu\psi\,\partial_\nu\psi),
\eea
while the second term gives a new contribution to the radion mass,
\bea
\delta  \mathcal{S}&\supset&-\frac{1}{2} \frac{4e^{4kL}\phi^2}{(1-v_\ir)^2}\int d^4x\, \psi^2.
\eea
Let us  focus on the kinetic term first, taking the limit $\xi_m\ll1$. Combining  the new kinetic term with the pre-existing kinetic terms gives:
\bea
\mathcal{S} +\delta \mathcal{S}&\supset&\int d^4x \, \left(
  -\mathcal{N}' \frac12 \eta^{\mu\nu}\partial_\mu\psi\partial_\nu\psi
  -\mathcal{N} \frac12 m^2 \psi^2 \right)  \:,
\eea
where the normalization factor is now
\bea
\mathcal{N}' &=&
  \frac{3M_*^3}{k}e^{2kL}
    \left(\frac{1}{1-v_\ir}+\frac{8\xi_\partial}{(1-v_\ir)^2}-\frac{e^{-2kL}}{1+v_\uv}\right)
    + \mathcal{O}(l^2) \nonumber\\
&=& \mathcal{N} +\frac{3M_*^3}{k}e^{2kL}
   \frac{8\xi_\partial}{(1-v_\ir)^2}+ \mathcal{O}(l^2) \:.
\label{massive_radion_norm_leading_order}
\eea
Interestingly, the new contribution to the kinetic term can apparently  cure the problem of a ghost-like radion for large IR curvature, provided one has
\bea
(v_\ir-1)&<& 8\xi_\partial.\label{eq:IR_kinetic_constraint}
\eea
Thus, for values of $v_\ir\approx15$, which can achieve a 750~GeV KK graviton, one obtains the constraint $\xi_\partial > 14/8=1.75$. For  our parametrization of the IR kinetic term in Eq.~\eqref{eq:new_ir_terms_rad}, it appears possible to avoid a ghost-like radion with $\xi_\partial =\mathcal{O}(1)$.
Note that the radion mass is now $m_r^2 = (\mathcal{N}/ \mathcal{N}') m^2$, which is non-tachyonic for the parameter space that avoids a ghost radion: the product $\mathcal{N}\times m^2$ is positive for $v_\ir>1$ (both $\mathcal{N}$ and $m^2$ are negative for $v_\ir>1$). Consequently, provided $\xi_\partial$ satisfies Eq.~\eqref{eq:IR_kinetic_constraint}, one has $\mathcal{N}'>0$ to ensure the radion kinetic term is positive definite and $m^2_r>0$ is automatically positive. Thus, the IR kinetic term for the radion in Eq.~\eqref{eq:new_ir_terms_rad}, which is consistent with the 4D symmetries of the theory, may help avoid the ghost-like radion that occurs for large values of $v_\ir$. 

With this observation we can reconsider the radion coupling to IR matter to include the effects of the IR localized kinetic term. We find that the IR coupling is modified to take the form:
\bea
\Lambda_\ir^{-1} &\simeq& \Lambda_{RS,\ir}^{-1}\times\sqrt{\frac{1+v_\uv}{v_\ir-1}} \left[\frac{8\xi_\partial}{(v_\ir-1)} -1\right]^{-1/2},
\eea
where we write the result for the case of $v_\ir>1$, assuming $\xi_\partial$ is chosen to ensure positivity of the radion kinetic term. The key point here is that avoiding the ghost-like radion has produced a brane coupling that is also well-behaved for $v_\ir>1$.

Turning now to the IR localized mass term, in the limit where the new kinetic piece is negligible, $\xi_\partial\ll1$, the quadratic action for the radion is 
\bea
\mathcal{S} &\supset&\mathcal{N} \int d^4x \, \left(
  -\frac12 \eta^{\mu\nu}\partial_\mu\psi\partial_\nu\psi
  -\frac12 (m^2+\delta m^2) \psi^2 \right)  \:,
\eea
where $\mathcal{N}$ is the prior normalization factor and the new mass correction from the localized IR  action is
\bea \label{eq:radion_mass_pert}
\delta m^2&\simeq&\frac{64\,\xi_m}{3(1-v_\ir)}\,l^2\,k^2e^{-2(k+u)L},
\eea
The mass $m^2$ was found earlier in Eq.~\eqref{radion_mass}. Observe that the mass correction has the same parametric dependence on the warp factor, the backreaction, and the IR curvature as $m^2$, namely $\delta m^2 \propto (1-v_\ir)^{-1} l^2 e^{-2(k+u)L}$.

At the end of Section \ref{sec:GW} we saw that the IR brane curvature term could not be used to significantly enhance the radion mass without making the radion interactions strongly coupled.  In this regard, it is interesting to note the effects of the brane mass term for $h_{55}$.  Working in the limit of small back reaction, $u\ll k$, and comparing Eq.~\eqref{eq:radion_mass_pert} to Eq.~\eqref{radion_mass}, we see that $\delta m^2/m^2\sim8k^2\xi_m/u^2$, seemingly allowing one to increase the mass.  Of course, if the boundary mass becomes too large one should incorporate it into the full BCs.

Let us emphasize that our comments in this section, regarding additional IR terms for $h_{55}$, should be understood as being mere speculations, motivated by the study of explicit brane masses for spin-2 metric fluctuations in Ref.~\cite{Chacko:2003yp}. Our comments generalize the approach of Ref.~\cite{Chacko:2003yp}  to include related terms for the radion. It is interesting, however, that such terms may offer some hope of avoiding a ghost-like radion. This observation may motivate more detailed study of the viability of these terms.
%%%%%%%%%%%%%%%%%%%%%%%%%%%%%%%%%%%%%%%%%%%%%%%%%%%%%%%%
\section{Comments on AdS/CFT\label{sec:ads/cft}}

According to the AdS/CFT correspondence~\cite{Maldacena:1997re}, RS models are
thought to be dual to strongly coupled 4D theories that are
(approximately) conformal for energies $M_*>E> e^{-kL}
M_*$~\cite{ArkaniHamed:2000ds}. Conformal symmetry is broken
explicitly in the UV by a cutoff (dual to the UV brane) and
spontaneously in the IR (dual to the IR brane). UV (IR) localized
fields in the 5D picture are dual to fundamental (composite) fields in
the 4D theory. More precisely, the UV value of a given bulk field in the 5D picture is dual to a fundamental field that is external to the strongly coupled 4D sector (see e.g.\ \cite{Batell:2007jv}). Here we make a few basic comments regarding RS models with brane curvature.\footnote{To the extent that the following discussion contains useful content, it is, in part, attributable to Ref.~\cite{agashe}. Any errors, however, are the responsibility of the authors.}

Recall that the effective 4D Planck mass   is
\bea
M_{Pl}^2=\frac{M_*^3v_\uv}{2k}+\frac{M_*^3}{2k}\left\{1 - (1-v_\ir)e^{-2kL} \right\},\label{eq:4D_planck_adscft}
\eea
including contributions from both the bulk and
brane curvatures. The different pieces have distinct interpretations in the dual 4D picture. The UV brane contribution, $M_\uv^2= v_\uv M_*^3/k$, results from a UV localized curvature term. As such, it corresponds to a kinetic term for the fundamental spin-2 particle associated with the UV restriction of the bulk 5D graviton~\cite{Agashe:2002jx}. The true massless graviton does not correspond exactly to this fundamental spin-2 field, but instead contains a small admixture of the massive spin-2 composite states. This admixture is tiny, however, as is evident by the high degree of UV localization for the massless zero-mode in the RS picture - that is, the UV value of the bulk graviton field is overwhelmingly dominated by the value of the zero mode (i.e.~massless graviton).  

The origin of this ``fundamental" contribution to the Planck mass is separate from the dual CFT dynamics.
 For $v_\ir\rightarrow0$, however, the remaining pieces in Eq.~\eqref{eq:4D_planck_adscft} encode a dynamically generated contribution to the Planck scale, induced by CFT loops; i.e., in the limit $M_\uv^2\rightarrow0$, the Planck mass (equivalently, massless graviton kinetic term) is fully induced by CFT loops. Taking the further limit $L\rightarrow\infty$, the RS expression for the Planck scale is $M_{Pl}^2\sim M_*^3/k$, which should correspond to the induced Planck mass from a CFT with UV cutoff $k$.  The latter has the form $M_{Pl}^2\sim c k^2$, with $c$ being uniquely determined by the corresponding central charge of the CFT. The holographic calculation of $c$ via 5D supergravity gives $(M_*/k)^3$~\cite{Henningson:1998gx}, so $M_{Pl}^2\sim M_*^3/k$ is in agreement with the RS result.\footnote{In the language of a dual large-$\mathcal{N}$ gauge theory, the induced Planck scale is $\sim k^2 \mathcal{N}^2$, where $\mathcal{N}^2\sim (M_*/k)^3$ relates to the number of colors in the dual CFT.} For finite $L$, the dual CFT has a further source of conformal symmetry breaking in the IR, at the scale $M_{IR}= e^{-kL} k$. Now the CFT-induced Planck mass is modified due to CFT symmetry breaking scale in the IR, giving $M_{Pl}^2\sim c(k^2-M_{IR}^2)$, in agreement with the limit $v_{\uv,\ir}\rightarrow 0$ of Eq.~\eqref{eq:4D_planck}. 

Turning on the IR term, $v_\ir\ne0$, the additional term in Eq.~\eqref{eq:4D_planck} encodes a change to the CFT-induced Planck mass due to some modification of the IR dynamics. While it is difficult to make precise statements about the strongly coupled sector in the dual 4D theory, it seems clear that the IR localized brane curvature is dual to some modification of the kinetic terms for the spin-2 composite states. Given that the massless graviton is largely comprised of the fundamental spin-2 field, one may not expect that modifying the composite spin-2 kinetic terms would affect the kinetic term for the massless graviton. However, the massless graviton contains a small admixture to the composite spin-2 states, and a modification to the kinetic terms for the spin-2 composites should induce a highly suppressed modification of the kinetic term for the massless graviton - i.e., it should generate a mixing-suppressed contribution to the Planck mass. This naive expectation is bourne out by Eq.~\eqref{eq:4D_planck_adscft}, where the suppressing factor $e^{-2kL}$ encodes the tiny mixing between the fundamental graviton and the spin-2 composites. Indeed, explicit calculations, in the so-called holographic basis, show that the mixing between the fundamental spin-2 state and the lightest composite spin-2 state is $\sin^2\theta_g\sim e^{-2kL}$~\cite{Batell:2007jv}, in agreement with the above.\footnote{Note that a massless mode from a bulk vector with IR kinetic term does not have this severe suppression of the IR-term dependence, as the fundamental/composite mixing is much larger in the spin-1 sector.}

Based on an inspection of the 4D Planck mass in Eq.~\eqref{eq:4D_planck_adscft}, one may naively interpret the effect of the IR term as corresponding to a change in the effective IR scale of the broken CFT. It is instructive to consider this point. The standard expression for the Planck mass in RS models, without brane curvature terms, can be written as
\bea 
M_{Pl}^2= \left(\frac{M_*}{2k}\right)^3 \times (k^2-M_{IR}^2).\label{eq:RSPlanck}
\eea
If one shifts the IR brane to a new location, $L\rightarrow L+\delta L$, the IR scale shifts accordingly to $M_{IR}'=e^{-k(L+\delta L)} k$ , modifying the expression for the Planck scale,
\bea 
M_{Pl}^2= \left(\frac{M_*}{2k}\right)^3 \times (k^2-(M_{IR}')^2).
\eea
Comparing this expression to Eq.~\eqref{eq:4D_planck_adscft}, it appears that the same effect can be obtained by including an IR brane curvature term with coefficient $v_\ir$, while keeping the brane fixed at $y=L$. Specifically, for $v_\ir<1$ we define
\bea
M_{IR}^{eff}=\sqrt{1-v_\ir} \;e^{-kL}k = \sqrt{1-v_\ir} \;M_{IR},
\eea
such that the 4D Planck mass Eq.~\eqref{eq:4D_planck} is written as 
\bea 
M_{Pl}^2= \left(\frac{M_*}{2k}\right)^3 \times (k^2-(M_{IR}^{eff})^2),
\eea
where we take $v_\uv=0$  to focus on the effect of the IR term. Comparing with the standard RS result~\eqref{eq:RSPlanck}, it appears that the effect of the IR curvature term  is to modify the effective IR scale. In particular, values in the range $0<v_\ir<1$ tend to decrease the effective IR scale in a way that appears similar to the increase in length $L\rightarrow L+\delta L$ with $\delta L =(\frac{-1}{2k})\times \log (1-v_\ir)$.

If this were correct, one could immediately deduce some additional  consequences of the IR curvature. In RS models, the radion couples conformally to IR localized fields  as $(r/\Lambda_{RS})\,T $, where $T$ is the trace of the stress-energy tensor and $\Lambda_{RS}$ is a dimensionful coupling whose value is set by the IR scale, $\Lambda_{RS}\sim M_{IR}$. With this information, one can guess the effect  of the IR curvature term on the coupling of the radion to IR localized fields:
\bea
\Lambda_{RS}\sim M_{IR}&\rightarrow& \Lambda\ \sim\ M_{IR}^{eff}\ =\  \sqrt{1-v_\ir} \;M_{IR}.
\eea
Thus, values of $0<v_\ir<1$, which tend to decrease $M_{IR}^{eff}$, would tend to \emph{increase} the coupling of the radion to IR fields, as this goes like $\Lambda^{-1}\sim (M_{IR}^{eff})^{-1}$. Conversely, values of $v_\ir<0$ tend to decrease the strength with which the radion couples. In Section~\ref{sec:radion-matter-GI} we explicitly calculated the radion coupling to IR matter in the presence of IR curvature and obtained a result in agreement with this naive guess.\footnote{Note that the effective coupling for the non-minimal term $hvR_\ir$ also has the expected form based on the above reasoning, once the radion kinetic term is brought to canonical form.} It is interesting that the above interpretation of the IR term allows one to foreshadow our conclusions so easily. Similarly, the interpretation of the IR curvature term as modifying the effective IR scale in the gravity sector suggests that the KK graviton masses should decrease for  $0<v_\ir<1$, consistent with explicit calculations~\cite{Davoudiasl:2003zt}.

While the above line of reasoning may  have utility, one should refrain from  taking the interpretation  of a modification to the IR confinement scale too seriously. This is evidenced by the failure of the IR curvature term to modify the KK masses for other bulk fields; i.e., the KK decomposition of a bulk vector gives a spectrum that is insensitive to the presence of an IR curvature term, implying that the spin-1 composite spectrum is not  sensitive to this modification. Thus, the interpretation in terms of a change to the IR scale appears to be a mere coincidence - the IR curvature represents a change to the kinetic terms for the composite states, which affects the massless graviton kinetic term via mixing, in a way that \emph{mimics} the effect of a modification to the IR/confinement scale.

Regarding the radion, it is interesting to note that the IR curvature affects the graviton and radion kinetic terms in different ways. The radion is highly IR-localized and is dual to a dilaton that is overwhelmingly composite. This situation is opposite to that of the graviton. Thus, the IR curvature, which encodes a modification to the kinetic terms for the spin-2 and dilaton sectors, should induce an unsuppressed change to the dilaton kinetic term. This behaviour is seen already in Eq.~\eqref{eq:radion_kinetic_GI}. The radion kinetic term is highly sensitive to the IR curvature, whereas it is relatively insensitive to the UV curvature, opposite to the massless graviton. These different sensitivities of the radion and graviton to the IR and UV curvature are consistent with the dual picture.

%%%%%%%%%%%%%%%%%%%%%%%%%%%%%%%%%%%%%%%%%%%%%%%%%%%%%
\section{Conclusion\label{sec:conc}}
The most general Lagrangian for RS models includes brane localized curvature terms on both the UV and IR branes. These terms can modify the spectrum of KK gravitons, as studied recently in relation to the 750~GeV diphoton excess. The brane curvature also has consequences for the properties of the radion. In this work we investigated some of these properties for a general RS model, both with and without GW stabilization. We showed that the brane curvature can modify the radion mass and couplings. Furthermore, demanding a non-ghost-like radion gives a restriction on the allowed parameter space for the curvature terms. We investigated the effects of a non-minimal IR coupling with the SM Higgs to determine the parameter space consistent with a non-ghost-like radion. Our results generalize a number of expression in Ref.~\cite{Csaki:2000zn} to the case with non-zero brane curvature. Unfortunately the resulting modifications did not remove the ghost-radion  encountered for $\mathcal{O}(10)$ values  of the IR curvature. Motivated by Ref.~\cite{Chacko:2003yp}, we also considered an explicit IR localized kinetic term for the radion. This term, which should be considered as speculative, may offer hope for avoiding the ghost-radion.
%%%%%%%%%%%%%%%%%%%%%%%%%%%%%%%%%%%%%%%%%%%%%%%%5
\section*{Acknowledgements\label{sec:ackn}}
BMD was supported by the Science and Technology Facilities Council (UK), DG was self-funded, and KM was supported by the Australian Research Council. The authors thank A.~Ahmed and V.~Sanz for helpful discussions and K.~Agashe for clarifying insights.

%%%%%%%%%%%%%%%%%%%%%%%%%%%%%%%%%%%%%%%%%%%%%%%%%%%%%
%%%%%%%%%%%%%%%%%%%%%%%%%%%%%%%%%%%%%%%%%%%%%%%%%%%%%
\section*{Appendix}
\appendix
In our analysis we included the Higgs-induced IR curvature in the full equations of motion and boundary conditions, arriving at an action, to quadratic order in the radion, with the form
\bea
\mathcal{S} &\supset&\mathcal{N} \int d^4x \, \left(
  -\frac12 \eta^{\mu\nu}\partial_\mu\psi\partial_\nu\psi
  -\frac12 m^2 \psi^2 \right)  \:, \label{eq:radion_quadratic_app}
\eea
with normalization factor
\bea
\mathcal{N} =
  \frac{3M_*^3}{k}e^{2kL}
    \left(\frac{1}{1-(v_\ir +\xi v_H)}-\frac{e^{-2kL}}{1+v_\uv}\right)
    + \mathcal{O}(l^2) \:.
\label{nonminimal_radion_norm_leading_order:app}
\eea
In the limit $v_H\ll 1$, an expansion to $\mathcal{O}(v_H)$ gives
\bea
\mathcal{N} =
  \frac{3M_*^3}{k}e^{2kL}
    \left(\frac{1}{1-v_\ir }-\frac{e^{-2kL}}{1+v_\uv}\right)
    + \frac{3M_*^3}{k}e^{2kL}
    \frac{\xi v_H }{(1-v_\ir) } +\ldots\:.
\label{nonminimal_radion_norm_leading_order:app}
\eea
The $\mathcal{O}(v_H)$ piece of the radion kinetic term agrees with that obtained by treating the non-minimal coupling term $\sim \xi v^2R_\ir$ as a perturbation on the background obtained without the IR Higgs.

%%%%%%%%%%%%%%%%%%%%%%%%%%%%%%%%%%%%%%%%%%%%%%%%5
%%%%%%%%%%%%%%%%%%%%%%%%%%%%%%%%%%%%%%%%%%%%%%%%5

%%%%%%%%%%%%%%%%%%%%%%%%%%%%%%%%%%%%%%%%%%%%%%%%%%%

\end{document}